\documentclass[apj]{emulateapj}
\usepackage{lscape}
\usepackage{apjfonts}
\usepackage{graphicx}

\slugcomment{Revised after referee report}
\shorttitle{Massive YSOs in the Galactic Center}
\shortauthors{An et~al.}

\begin{document}
\title{First Spectroscopic Identification of Massive Young Stellar Objects\\
in the Galactic Center}

\author{Deokkeun An\altaffilmark{1},
Solange V.\ Ram\'irez\altaffilmark{1},
Kris Sellgren\altaffilmark{2},
Richard G. Arendt\altaffilmark{3,4},
A.\ C.\ Adwin Boogert\altaffilmark{1},\\
Mathias Schultheis\altaffilmark{5,6},
Susan R.\ Stolovy\altaffilmark{7},
Angela S.\ Cotera\altaffilmark{8},\\
Thomas P.\ Robitaille\altaffilmark{9,10}, and
Howard A.\ Smith\altaffilmark{9}
}

\altaffiltext{1}{Infrared Processing and Analysis Center,
California Institute of Technology, Mail Stop 100-22, Pasadena, CA 91125;
deokkeun@ipac.caltech.edu,solange@ipac.caltech.edu.}
\altaffiltext{2}{Department of Astronomy, Ohio State University,
140 West 18th Avenue, Columbus, OH 43210; sellgren@astronomy.ohio-state.edu.}
\altaffiltext{3}{CRESST/UMBC/GSFC, Code 665, NASA/Goddard Space Flight Center,
8800 Greenbelt Road, Greenbelt, MD 20771.}
\altaffiltext{4}{Science Systems and Applications, Inc.}
\altaffiltext{5}{Observatoire de Besan\c{c}on, 41bis, avenue de l'Observatoire,
F-25000 Besan\c{c}on, France.}
\altaffiltext{6}{Institut d'Astrophysique de Paris, CNRS, 98bis Bd Arago,
F-75014 Paris, France.}
\altaffiltext{7}{Spitzer Science Center, California Institute of Technology,
Mail Code 220-6, 1200 East California Boulevard, Pasadena, CA 91125.}
\altaffiltext{8}{SETI Institute, 515 North Whisman Road, Mountain View, CA 94043.}
\altaffiltext{9}{Harvard-Smithsonian Center for Astrophysics, 60 Garden Street,
Cambridge, MA 02138.}
\altaffiltext{10}{Spitzer Postdoctoral Fellow.}

\begin{abstract}
We report the detection of several molecular gas-phase and ice absorption features
in three photometrically-selected young stellar object (YSO) candidates in the central
280~pc of the Milky Way. Our spectra, obtained with the Infrared Spectrograph (IRS)
onboard the {\it Spitzer Space Telescope}, reveal gas-phase absorption from CO$_2$
($15.0\mu$m), C$_2$H$_2$ ($13.7\mu$m) and HCN ($14.0\mu$m). We attribute this absorption
to warm, dense gas in massive YSOs. We also detect strong and broad $15\mu$m CO$_2$
ice absorption features, with a remarkable double-peaked structure. The prominent
long-wavelength peak is due to CH$_3$OH-rich ice grains, and is similar to those found
in other known massive YSOs. Our IRS observations demonstrate the youth of these objects,
and provide the first spectroscopic identification of massive YSOs in the Galactic Center.
\end{abstract}

\keywords{infrared: ISM
--- ISM: molecules
--- stars: formation}

\section{Introduction}

The Central Molecular Zone (CMZ) is the innermost $\sim200$~pc region of the Milky
Way Galaxy. It is a giant molecular cloud complex delineated by a gradient in the
CO column density and temperature. The CMZ contains $\sim10\%$
of the Galaxy's molecular gas, and produces $5\%$--$10\%$ of its infrared and Lyman
continuum luminosities \citep[see a review by][and references therein]{morris:96}.

Evidence is mounting that conditions for star formation in the CMZ are significantly
different from those in the Galactic disk. The gas pressure and temperature are higher
in the CMZ than in the average disk, conditions that favor a larger Jeans mass for
star formation and an initial mass function biased towards more massive stars.
Furthermore, the presence of strong magnetic fields, tidal shear, and turbulence
challenges the standard paradigm of slow gravitational collapse of molecular cloud cores.

The CMZ provides several signposts of {\it in situ} star formation, such as H$_2$O
masers, (ultra-)compact \ion{H}{2} regions, young OB stars, and young supernova
remnants. However, young stellar objects (YSOs or protostars), which are the direct
tracers of current star formation, have so far eluded detection in the CMZ. They have
been inferred to be present based on infrared photometry \citep[e.g.,][]{felli:02,
schuller:06,yusefzadeh:09}, but spectroscopic observations are required to confirm
their status as a YSO. This is because evolved stars can look like YSOs in broad-band
photometry, if they are heavily dust attenuated \citep[e.g.,][]{schultheis:03}, a
problem towards the Galactic Center (GC), where $A_V \approx 30$.

In this {\it Letter}, we present spectroscopic follow-up observations of YSO
candidates in the CMZ, using the Infrared Spectrograph \citep[IRS;][]{houck:04}
onboard the {\it Spitzer Space Telescope} \citep{werner:04}. Massive YSO candidates
were photometrically selected from the point source catalog \citep{ramirez:08},
which was extracted from images of the CMZ \citep{stolovy:06} made using the
Infrared Array Camera \citep[IRAC;][]{fazio:04}. This high sensitivity and high
spatial resolution image has led to a better identification of YSO candidates
and their follow-up spectroscopic observations.

\section{Photometric Sample Selection}

The IRAC point source catalog \citep{ramirez:08} contains photometry for more than a
million point sources in the entire CMZ ($2\arcdeg \times 1.4\arcdeg$ or $280 \times
200$~pc) in four channels ($3.6\mu$m, $4.5\mu$m, $5.8\mu$m, and $8.0\mu$m). Initially,
we selected point sources with ${\rm [3.6] - [8.0]} \geq 2.0$, corresponding to
YSOs with $M_* \ga 2.5 M_\odot$ \citep{whitney:03,whitney:04}. We further confined the
sample to those within $|b| < 15\arcmin$, resulting in $1207$ objects. When we had
photometric measurements in at least 5 bandpasses from IRAC, 2MASS \citep[$JHK_s$;][]
{skrutskie:06}, and/or ISOGAL \citep[$7\mu$m and $15\mu$m;][]{omont:03}, we selected
YSO candidates by comparing the observed spectral energy distribution (SED) with YSO
models \citep{robitaille:06} using a SED fitting tool by \citet{robitaille:07}.
Otherwise, we applied additional color constraints from \citet[][${\rm [3.6]-[4.5]}
\geq 0.5$, ${\rm [4.5]-[5.8]} \geq 0.5$, and ${\rm [5.8]-[8.0]} \geq 1.0$]{whitney:04}
to identify YSO candidates. SED fitting and color selection narrowed down our sample
to about $200$ objects.

Then, we carefully inspected IRAC three-color images to select objects that are
distinct within the IRS slit entrances against the crowded stellar field and bright
local background. Finally, a literature search was carried out for the selected objects,
and one Wolf-Rayet star and four OH/IR stars were discarded. Our final sample is composed
of 107 objects, among which 25 were previously known YSO candidates from ISOGAL
\citep{felli:02}.

\section{IRS Observations and Data Reduction}

\begin{figure*}
\epsscale{1.0}
\plotone{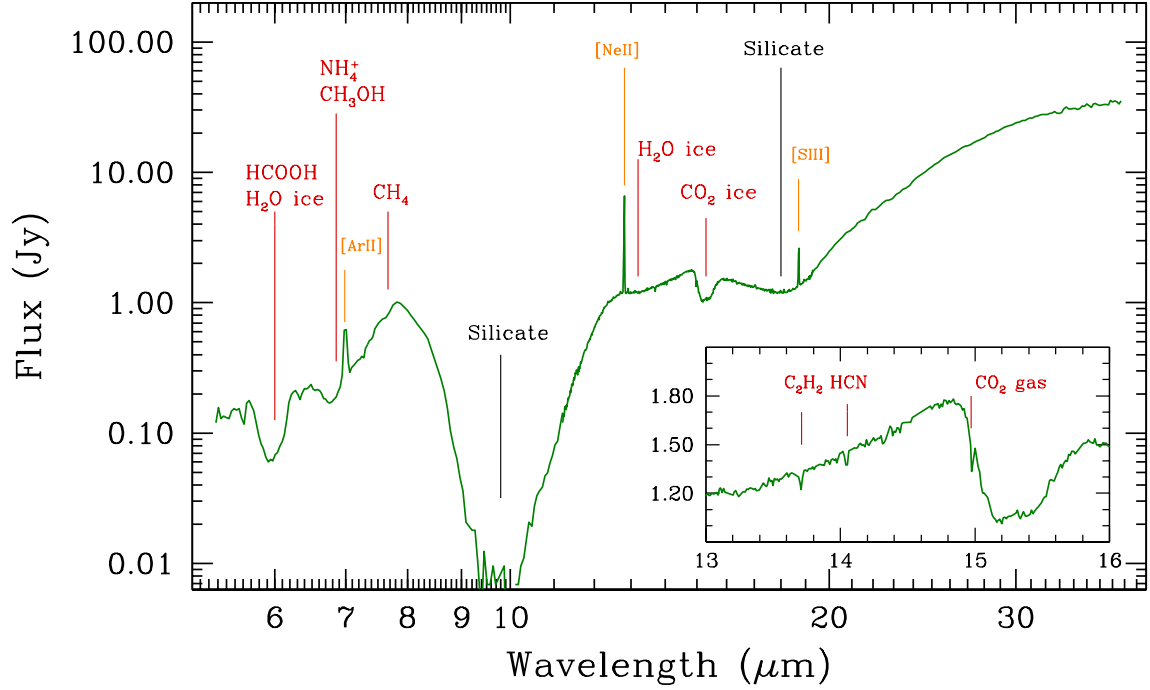}
\caption{Composite IRS spectrum of SSTGC~797384. The spectrum is from SL at
$\lambda \le 11.2\mu$m, SH at $11.2\mu$m $\leq \lambda \leq 19.3\mu$m, and LL at
$\lambda \geq 19.3\mu$m. This composite spectrum is characterized by an extremely
red spectral energy distribution, strong and deep silicate absorption, and
several molecular gas- and solid-phase absorptions.
\label{fig:all}}
\end{figure*}

\begin{deluxetable*}{lcccc}
\tablewidth{0pt}
\tablecaption{Properties of the Sample\label{tab:tab1}}
\tablehead{
  \colhead{Quantities} &
  \colhead{Units} &
  \colhead{SSTGC~524665} &
  \colhead{SSTGC~797384} &
  \colhead{SSTGC~803187}
}
\startdata
R.A.(J2000.0)                        & h:m:s              & 17:45:39.86    & 17:47:23.68    & 17:47:26.29    \nl
Decl.(J2000.0)                       & d:m:s              & -29:23:23.4    & -28:23:34.6    & -28:22:1.5     \nl
UKIDSS $J$\tablenotemark{a}          & mag                & \nodata        & $18.23\pm0.06$ & $17.39\pm0.03$ \nl
UKIDSS $H$\tablenotemark{a}          & mag                & \nodata        & $14.68\pm0.01$ & $16.60\pm0.05$ \nl
UKIDSS $K$\tablenotemark{a}          & mag                & $15.71\pm0.10$ & $12.92\pm0.01$ & $14.37\pm0.02$ \nl
IRAC ${\rm[3.6]}$\tablenotemark{b}   & mag                & $11.42\pm0.01$ & \nodata        & $12.22\pm0.02$ \nl
IRAC ${\rm[4.5]}$\tablenotemark{b}   & mag                & $ 8.63\pm0.01$ & $9.41\pm0.01$  & $ 8.97\pm0.01$ \nl
IRAC ${\rm[5.8]}$\tablenotemark{b}   & mag                & $ 7.08\pm0.01$ & $7.66\pm0.01$  & $ 7.24\pm0.01$ \nl
IRAC ${\rm[8.0]}$\tablenotemark{b}   & mag                & $ 6.13\pm0.01$ & $5.64\pm0.01$  & $ 5.11\pm0.01$ \nl
MIPS ${\rm[24]}$\tablenotemark{c}    & mag                & $ 1.54\pm0.01$ & $0.55\pm0.01$  & \nodata        \nl
${\rm T_{\rm ex}~(C_2H_2~gas)}$      & K                  & $300\pm150$    & $200\pm150$    & $300\pm150$    \nl
${\rm T_{\rm ex}~(HCN~gas)}$         & K                  & $400\pm50$     & $100\pm50$     & $100\pm50$     \nl
${\rm T_{\rm ex}~(CO_2~gas)}$        & K                  & $200\pm50$     & $100\pm50$     & $100\pm50$     \nl
${\rm N_{col}~(C_2H_2~gas)}$         & $10^{16}$cm$^{-2}$ & $7.9\pm3.3$    & $1.0\pm0.3$    & $2.0\pm0.8$    \nl
${\rm N_{col}~(HCN~gas)}$            & $10^{16}$cm$^{-2}$ & $15.8\pm4.6$   & $1.0\pm0.4$    & $2.0\pm0.9$    \nl
${\rm N_{col}~(CO_2~gas)}$           & $10^{16}$cm$^{-2}$ & $20.0\pm5.6$   & $5.0\pm1.5$    & $7.9\pm2.8$    \nl
${\rm N_{col}~(CO_2~solid)}$         & $10^{19}$cm$^{-2}$ & $0.11\pm0.01$  & $0.13\pm0.01$  & $0.21\pm0.01$  \nl
${\rm N_{col}~(H_2O~solid, 6\mu m)}$ & $10^{19}$cm$^{-2}$ & $<1.7$         & $<2.3$         & $<4.7$         \nl
${\rm N_{col}~(H_2O~solid, 13\mu m)}$& $10^{19}$cm$^{-2}$ & $0.6\pm0.4$    & $1.3\pm0.4$    & $1.9\pm0.3$    \nl
${\rm N_{col}~(H_2)}$                & $10^{22}$cm$^{-2}$ & $2.3\pm0.2$    & $5.0\pm0.5$    & $5.8\pm0.7$    \nl
${\rm N_{C_2H_2} / N_{H_2}}$         & $10^{-7}$          & $34.3\pm14.7$  & $2.0\pm0.6$    & $3.4\pm1.4$    \nl
${\rm N_{HCN} / N_{H_2}}$            & $10^{-7}$          & $68.7\pm20.9$  & $2.0\pm0.8$    & $3.4\pm1.6$    \nl
${\rm N_{CO_2,gas} / N_{H_2}}$       & $10^{-7}$          & $87.0\pm25.5$  & $10.0\pm3.2$   & $13.6\pm5.1$   \nl
${\rm N_{CO_2,gas} / N_{CO_2,solid}}$&                    & $0.18\pm0.05$  & $0.04\pm0.01$  & $0.04\pm0.01$  \nl
${\rm N_{CO_2,solid} / N_{H_2O,solid}}$&                  & $0.18\pm0.12$  & $0.10\pm0.03$  & $0.11\pm0.02$  \nl
$A_V~(\tau_{9.7\mu m})$              & mag                & $27\pm3$       & $53\pm5$       & $62\pm7$       \nl
$A_V$~(color)\tablenotemark{d}       & mag                & $29.0\pm2.6$   & $32.0\pm2.6$   & $27.0\pm5.4$   
\enddata
\tablecomments{Random errors are shown in the photometry.}
\tablenotetext{a}{Aperture3 magnitudes from UKIDSS DR2 \citep{warren:07}.}
\tablenotetext{b}{Photometry from \citet{ramirez:08}.}
\tablenotetext{c}{Photometry from MIPSGAL (S.\ Carey, 2008, private communication).}
\tablenotetext{d}{Based on the 2MASS and IRAC color-magnitude diagrams of GC red
giant branch stars within $2\arcmin$ of the source \citep{schultheis:09}.}
\end{deluxetable*}

We obtained spectroscopic data for 107 YSO candidates using the four IRS modules
in May and October 2008. We observed each target in IRS staring mode with 4 exposures
per source (2 cycles). Exposure times were
6~sec--120~sec in SH (short-high; short wavelength, high resolution), 6~sec--60~sec
in LH (long-high), 6~sec--14~sec in SL (short-low), and 6~sec in LL (long low) modules,
depending on the source's brightness,
to achieve a signal-to-noise ratio (S/N) of at least $50$ in SH and SL, and a minimum
S/N of $10$ in LH and LL. We reduced the IRS spectra from the basic calibrated data (BCD)
products version S17.2.0 and S18.1.0, using the SSC software packages {\tt IRSCLEAN}
(to correct for bad pixel values) and {\tt SPICE} (to extract spectra).

Because the GC exhibits strong, spatially variable background, we observed multiple 
off-source measurements (one cycle, $1\times1$ mapping mode) to derive backgrounds 
near each of our YSO candidates in the four IRS modules.
The on-source and the off-source observations were taken consecutively to minimize 
zodiacal light and instrumental variations.
For the high resolution observations, we observed and extracted four background 
positions ($\sim \pm1\arcmin$ offsets in either R.A.\ or Decl.).
For the low resolution observations, we took spectra from two background positions 
at $\sim \pm1\arcmin$ away in the direction perpendicular to the slit, and extracted
two additional background spectra at positions along the on-source slit.
In all of the four different IRS modules, we tried to extract the background spectra 
at the same position as much as possible, to minimize the flux difference from
different modules. 

We made an interpolation of a plane in three dimensional space (positions on the
IRAC map and wavelength) to obtain a background spectrum at the source position.
We estimated an error in each source's background from the dispersion of four
different background spectra, constructed from alternate sets of three out of the
four background pointings.

A complete analysis of spectra for all of our 107 YSO candidates will be presented
elsewhere (D.\ An et al. 2009, in preparation). For the current analysis, we selected
three targets (Table~\ref{tab:tab1}) from among those showing characteristic spectral
features of massive YSOs, which include gaseous molecular absorptions from C$_2$H$_2$,
HCN, CO$_2$ \citep[e.g.,][]{lahuis:00,boonman:03,knez:09}, and a solid-phase absorption
from CO$_2$ ice bending mode \citep[e.g.,][]{gerakines:99}.

Both SSTGC~797384 and SSTGC~803187 are associated with a relatively weak radio
continuum source \citep[SGR~B2(P) and SGR~B2(R), respectively;][]{mehringer:93}.
They are on the outskirts of the Sgr~B2 molecular cloud ($\sim2$~pc--$4$~pc from
the well-studied radio source SGR~B2(M)), which is one of the most active complexes
of compact \ion{H}{2} regions in the Galaxy \citep[e.g.,][]{mehringer:95}.
\citet{mehringer:93} derived zero-age main-sequence spectral types of B0 and O6.5
for these compact \ion{H}{2} regions, respectively, from the number of ionizing
photons. SSTGC~524665 does not have radio continuum emission associated with it.
However, it is coincident with an H$_2$O maser \citep{forster:89}, and is
adjacent to a region of $4.5\mu$m excess emission \citep{yusefzadeh:09}, possibly
tracing shocked molecular outflows \citep[e.g.,][]{smith:06}.

For SSTGC~803187, we used a non-standard extraction aperture in SL, because of a
nearby source ($\approx7\arcsec$ south of the target) along the slit.  We followed
the prescription on the IRS data reduction website\footnote{See
http://ssc.spitzer.caltech.edu/IRS/calib.} to calibrate the flux. We trimmed the
end of the orders to remove the noisy part of spectra, and spectra from different
orders in high-resolution modules were averaged using a linear ramp. After background
subtraction, the SH and LH spectra were scaled down in flux
to LL over the common wavelength interval for SSTGC~797384 and SSTGC~803187. The SL
spectra were then scaled to SH. For these sources, we assumed that the flux mismatch
is due to narrower slit entrances in SH and SL. For SSTGC~524665, we used the SL as
a basis for the scaling, because our observations in LL and LH were contaminated by
extended emission from a nearby ($\approx10\arcsec$ southwest of the target)
bright source on the $24\mu$m MIPS image \citep{carey:09,yusefzadeh:09}. The background
for this target is likely to be over-subtracted, because the target lies on a dark
cloud with high extinction, while background spectra were taken at brighter spots.
The potential problem of the background subtraction results in H$_2$ lines (arising
from the surrounding sky) appearing in absorption in SSTGC~524665. In the following
initial analysis, we did not use LH data for all targets, but focused on the spectral
features in other modules.

\section{Analysis and Results}

Figure~\ref{fig:all} displays background-subtracted spectra of SSTGC~797384, in
SL ($\lambda \leq 11.2\mu$m), SH ($11.2\mu$m $\leq \lambda \leq 19.3\mu$m), and LL
($\lambda \geq 19.3\mu$m). The observed spectrum is characterized by an extremely red
SED [$\alpha \equiv d\log(\lambda F_\lambda)/d\log(\lambda) \approx 2$], strong and
deep silicate absorptions at $9.7\mu$m and $18\mu$m, ice absorption features at
$6\mu$m, $6.85\mu$m, $13\mu$m, and $15.2\mu$m. Although the presence of forbidden
lines indicates that these objects are likely associated with an (ultra-)compact
\ion{H}{2} region, it could be also due to under-subtracted emissions from the
background.

\begin{figure}
\epsscale{1.15}
\plotone{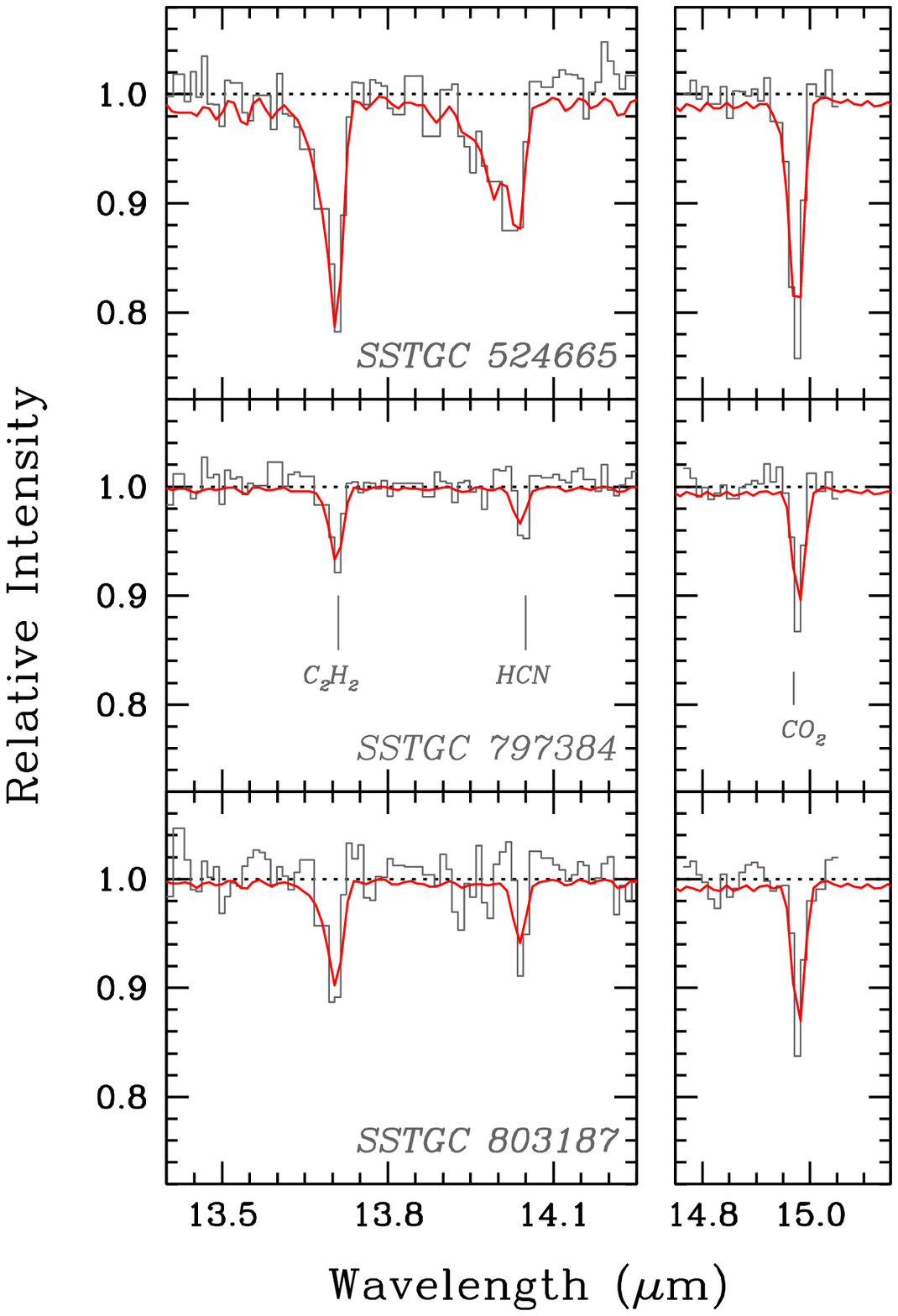}
\caption{Gas-phase molecular absorptions from C$_2$H$_2$ $\nu_5 = 1-0$ ($13.71\mu$m),
HCN $\nu_2 = 1-0$ ($14.05\mu$m), and CO$_2$ $\nu_2 = 1-0$ ($14.97\mu$m). Best-fitting
models are shown in solid lines.
\label{fig:gas}}
\end{figure}

\begin{figure}
\epsscale{1.15}
\plotone{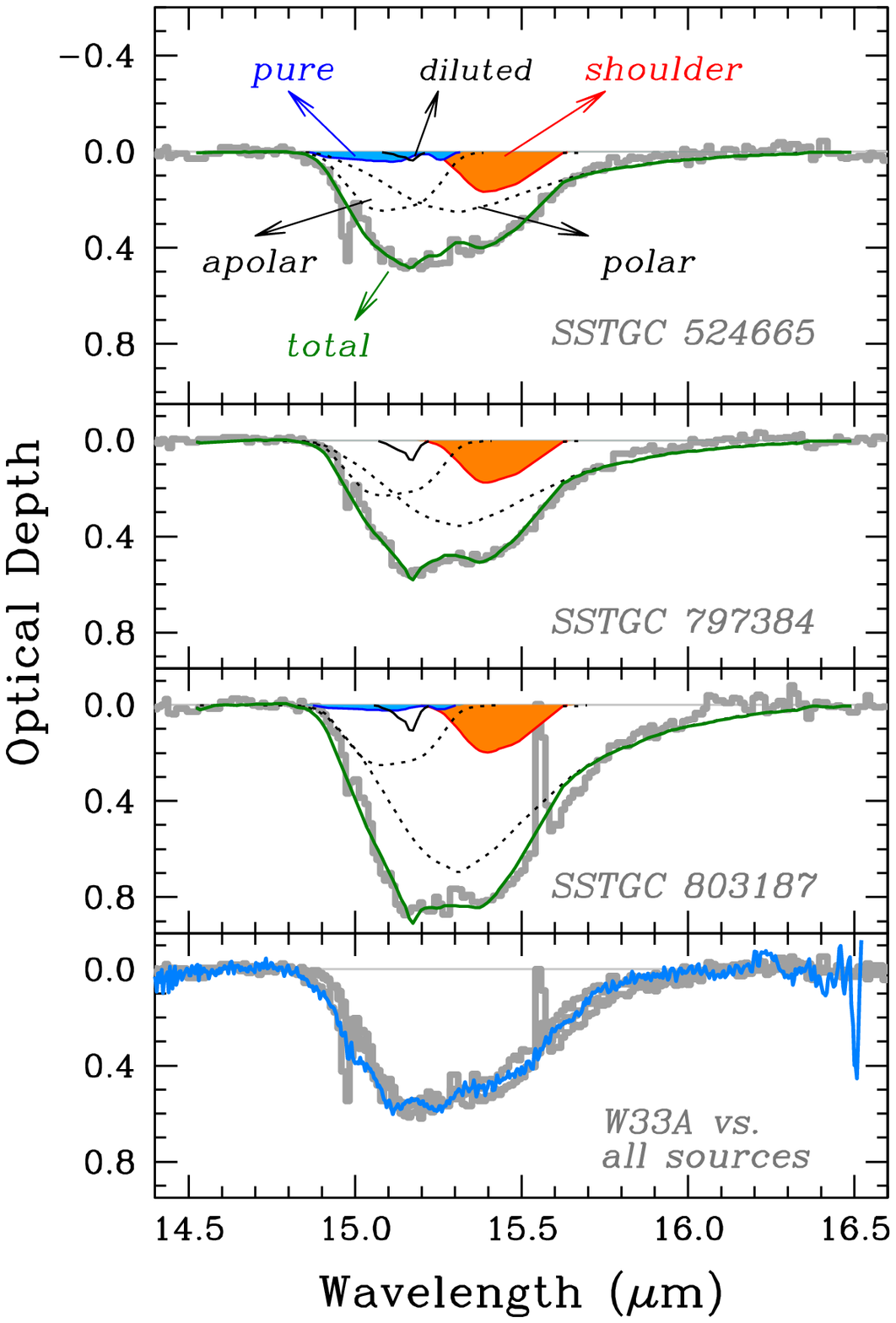}
\caption{Optical depth spectra of solid-phase absorption from the CO$_2$ ice bending
mode. Best-fitting CO$_2$ ice models and individual CO$_2$ ice components are shown
for each target: polar (dotted line, centered at $\sim15.3\mu$m), apolar (dotted line,
centered at $\sim15.1\mu$m), pure (blue shaded), diluted (black solid line), $15.4\mu$m
shoulder (orange-shaded), and the sum of these absorption components (green line). The
bottom panel shows a comparison of the ice absorption profile between our sources (grey)
and massive YSO W33A (blue). The optical depths for our targets were scaled in the bottom
panel for comparison.
\label{fig:co2}}
\end{figure}

Figure~\ref{fig:gas} shows gas-phase molecular absorptions at $13.71\mu$m (C$_2$H$_2$
$\nu_5 = 1-0$), $14.05\mu$m (HCN $\nu_2 = 1-0$), and $14.97\mu$m (CO$_2$ $\nu_2 = 1-0$),
detected in three YSO candidates.
To derive the excitation temperature ($T_{\rm ex}$) and column density ($N_{\rm col}$)
for each molecular species, we used model spectra from \citet{spectrafactory}
based on {\tt HITRAN04} linelist \citep{hitran} for C$_2$H$_2$ and HCN, and those based
on {\tt HITEMP} \citep{hitemp} for CO$_2$. A second order polynomial was used to set a
local continuum at $13.30\mu$m $\leq \lambda \leq 14.55\mu$m for C$_2$H$_2$ and HCN, and
$14.77\mu$m $\leq \lambda \leq 15.06\mu$m for CO$_2$. We did not include isotopes in the
computation because of the limited parameter span in the model grids.  However, even a
relatively high fraction of isotopes in GC \citep[$^{12}{\rm C}/^{13}{\rm C}\approx23$;]
[]{wannier:80} has a negligible impact in the model fitting.

We first made a fit to C$_2$H$_2$, and subtracted its contribution to the absorption
near weaker HCN bands. Best-fitting model $T_{\rm ex}$ and $N_{\rm col}$ were found by
searching the minimum $\chi^2$ of the fits over 100~K $\leq T_{\rm ex} \leq$ 1000~K in
steps of $\Delta T_{\rm ex} = 100$~K, and $15 \leq \log{N_{\rm col}} \leq 18$ for C$_2$H$_2$,
$16 \leq \log{N_{\rm col}} \leq 18$ for HCN, and $16 \leq \log{N_{\rm col}} \leq 22$ for
CO$_2$ with intervals of $0.1$~dex. Solid lines in Figure~\ref{fig:gas} show our
best-fitting models, and their $T_{\rm ex}$ and $N_{\rm col}$ are listed in
Table~\ref{tab:tab1}. Errors in these parameters were estimated from $\Delta \chi^2$,
where $1\sigma$ measurement errors were taken from the scatter of flux in the spectra.
Systematic errors from background subtraction and nodding differences were then added
in quadrature. We tested with varying covering factors, but found that best-fitting
case yields its value equal to or close to unity.

These gaseous bandheads have been detected in absorption toward YSOs, tracing
the warm and dense gas in the circumstellar disk and/or envelopes
\citep[e.g.,][]{lahuis:00,boonman:03,knez:09}. They are sometimes detected in
the photosphere and/or the circumstellar envelope of carbon-rich asymptotic giant branch
stars \citep[e.g.][]{aoki:99}, but carbon stars have not been found in the GC region
\citep[e.g.,][]{guglielmo:98}.

The above estimates are based on models with a Doppler parameter $b = 3\ {\rm km\ s^{-1}}$.
The line width measurements of these molecules for several massive YSOs and that
of the strongest H$_2$CO absorption components near SSTGC~803187 are in the range of
$b = 1-7\ {\rm km\ s^{-1}}$ \citep[e.g.,][]{mehringer:95,vandertak:00,knez:09}.
There are limited model grids at $b = 10\ {\rm km\ s^{-1}}$ for C$_2$H$_2$ and HCN, but
$T_{\rm ex}$ and $N_{\rm col}$ were generally found within $2\sigma$ from those at
$b = 3\ {\rm km\ s^{-1}}$.

\begin{figure}
\epsscale{1.15}
\plotone{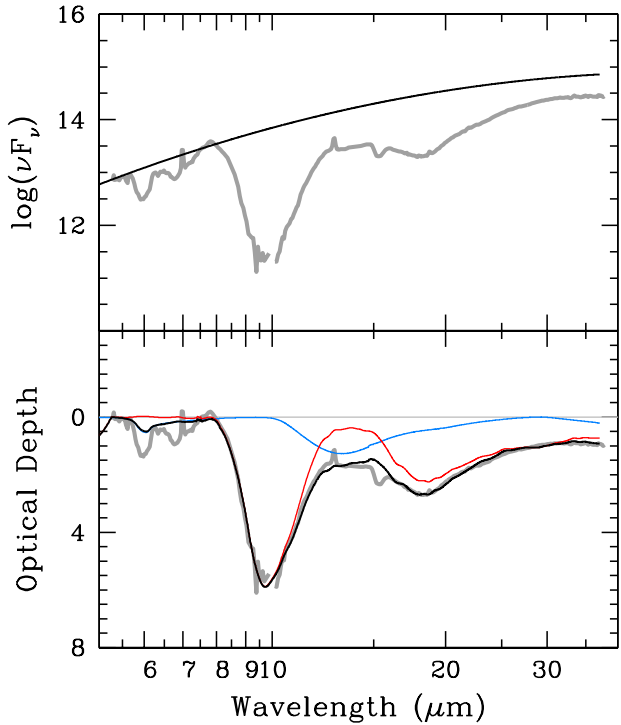}
\caption{Fit to the H$_2$O ice and silicate absorption for SSTGC~797384.
{\it Top:} SL and LL data (grey), with a best-fitting pseudo-continuum (black line).
{\it Bottom:} Decomposition of optical depth spectra (grey) with the silicate
(red) and the laboratory H$_2$O ice profiles (blue). Black line represents a sum
of these two components.
\label{fig:continuum}}
\end{figure}

Figure~\ref{fig:co2} shows optical depth spectra of our sources (grey) at $\sim15.2\mu$m,
where the strong and wide CO$_2$ ice absorption is seen.  We set a local continuum over
$14.5\mu$m $\leq \lambda \leq 16.5\mu$m using a 3rd order polynomial, and followed the
prescription in \citet{pontoppidan:08} to decompose the absorption profile with five
laboratory spectral components: polar (CO$_2$:H$_2$O $= 14:100$ at 10~K; dotted line,
centered at $\sim15.3\mu$m), apolar (CO:CO$_2 = 100:70$ at 10~K; dotted line, centered
at $\sim15.1\mu$m), pure CO$_2$ (15~K; blue shaded), diluted CO$_2$ (CO:CO$_2 = 100:4$
at 10~K; black solid line), and $15.4\mu$m shoulder CO$_2$ ice profile (modeled with two
Gaussians in wavenumber space; orange shaded). We found a best-fitting set of models from
the non-linear least squares fitting routine MPFIT \citep{markwardt:09}. Green solid line
represents the sum of all of the ice components, and the CO$_2$ ice column density in
Table~\ref{tab:tab1} was estimated from the integrated absorption, adopting the integrated
line strength $A = 1.1\times10^{-17} {\rm cm\ molecule^{-1}}$ \citep{gerakines:95}.

Unlike the CO$_2$ absorption profiles observed in quiescent molecular clouds
\citep[e.g.][]{whittet:09}, the $15.2\mu$m band in Figure~\ref{fig:co2} shows a remarkable
double-peaked profile. Double peaked profiles are commonly observed toward YSOs
\citep[e.g.,][]{gerakines:99,pontoppidan:08}, and are ascribed to pure CO$_2$ ices
resulting from crystallization of heated H$_2$O-rich ices. However, the double peaks toward
the GC candidate YSOs are centered at longer wavelengths ($15.15\mu$m and $15.4\mu$m
vs.\ $15.10\mu$m and $15.25\mu$m), and result from CO-rich ($15.15\mu$m peak) and
CH$_3$OH-rich ices ($15.4\mu$m peak; see Fig.~\ref{fig:co2}).

The strength of the $15.4\mu$m peak is similar to that of the well-studied embedded
massive YSO W33A \citep[][bottom panel in Fig.~\ref{fig:co2}]{gerakines:99}. It is
ascribed to a Lewis acid-base interaction of CO$_2$ (the Lewis acid) with CH$_3$OH
\citep{dartois:99a}. Other species could be acting as a base as well, but CH$_3$OH is
preferred due to its high abundance toward W33A: $5\%$--$22\%$ relative to solid H$_2$O
\citep{dartois:99b}. Two other YSOs (AFGL~7009S, AFGL~2136) show a prominent $15.4\mu$m
peak, and indeed these sources have high CH$_3$OH abundances as well
\citep{dartois:99b,gibb:04}.
This suggests that the GC candidate YSOs have high solid CH$_3$OH abundances as
well.\footnote{This needs to be verified by independent L-band spectroscopy of the
$3.53\mu$m C-H stretch mode of CH$_3$OH \citep[e.g.,][]{dartois:99b}.}
Although the origin of the large quantities of CH$_3$OH in
the previously studied massive YSOs is not fully understood \citep{dartois:99a},
so far all lines of sight with high solid CH$_3$OH abundances are associated with star
formation, strengthening the idea that the sources studied in this paper are indeed YSOs.

To derive abundances of these molecular absorptions with respect to the hydrogen
and solid H$_2$O column densities, we followed the procedure in \citet{boogert:08}
to fit the H$_2$O ice and silicate absorption profiles to SL and LL spectra.
Figure~\ref{fig:continuum} shows an example for SSTGC~797384. We used the silicate
absorption profiles in the line of sight to the GC \citep[GCS~3 spectrum;][]{kemper:04}
plus a laboratory spectrum of pure amorphous H$_2$O ice at $T = 10$~K \citep{hudgins:93}.
We simultaneously fit a second-order polynomial for a pseudo-continuum (i.e., including
corrections for the continuous extinction), the silicate profile, and H$_2$O ice
absorption to the $5\mu$m $\leq \lambda \leq 32\mu$m spectrum. We masked absorption
features at $6\mu$m, $7\mu$m, and $15\mu$m, and all unresolved emission lines, before
performing a non-linear least squares fit.

Best-fitting parameters are listed in Table~\ref{tab:tab1}. We obtained a total
hydrogen column density from the optical depth of the $9.7\mu$m silicate absorption,
assuming $A_V / \tau_{9.7} = 9$ \citep{roche:85} and $N_{\rm H} / A_V \approx 1.87
\times 10^{21}$ cm$^{-2} {\rm mag}^{-1}$ \citep{bohlin:78} at $R_V = 3.1$. The H$_2$
column density was then approximated by $N_{\rm H_2} = N_{\rm H}/2$. The ice column
density for the $13\mu$m librational H$_2$O absorption was estimated from the
integrated absorption of the best-fitting H$_2$O model. The H$_2$O ice column density
from the $6\mu$m bending mode, fit separately after fixing the continuum and extinction
to previously found values, is an upper limit because the $6\mu$m absorption is not
solely due to H$_2$O ice.
We adopted the integrated line strengths $A = 1.2\times10^{-17} {\rm cm\ molecule^{-1}}$
for the bending mode and $A = 3.1\times10^{-17} {\rm cm\ molecule^{-1}}$ for the
librational mode \citep{gerakines:95}. Errors in these parameters (Table~\ref{tab:tab1})
are formal estimates made by varying the range of wavelengths that we used for the
$9.7\mu$m silicate fitting, or by taking a few different ways of setting the continuum.

The gas-phase molecular abundances relative to H$_2$ are listed in Table~\ref{tab:tab1}.
Our derived abundances of $\sim10^{-7}$--$10^{-6}$ for C$_2$H$_2$ and HCN are comparable
to those found for massive YSOs \citep{lahuis:00,knez:09}, although abundances for
SSTGC~524665 have large errors. Intervening molecular clouds in the line of sight
to the GC are less likely the main cause of these absorptions, because the average
HCN abundance of $2.5\times10^{-8}$ \citep{greaves:96} towards Sgr~B2(M) is
an order of magnitude lower than our measurements. Our gas-phase CO$_2$ abundances
are an order of magnitude larger than those found towards massive YSOs in \citet{boonman:03},
but our gas to solid abundance ratios for CO$_2$ are consistent with their estimates
($10^{-1}$--$10^{-2}$). Our abundance of CO$_2$ ice relative to H$_2$O ice is within
the range ($0.10$--$0.23$) found towards massive YSOs \citep{gerakines:99}.

Finally, Table~\ref{tab:tab1} lists our estimates on $A_V$ from the $9.7\mu$m silicate
absorption and those from \citet{schultheis:09}, based on the 2MASS and IRAC
color-magnitude diagrams of GC red giant branch stars within $2\arcmin$ of the
source. Both SSTGC~797384 and SSTGC~803187 have higher $A_V$ values than the average
for field stars, implying that a significant fraction of the attenuation is
intrinsic to the source. SSTGC~524665 has a lower $A_V$, comparable to the average
value for surrounding field stars.
We also note that SSTGC~524665 is located at $b \approx -0.2\arcdeg$, so it is possible
that it is in front of the GC.  If we assume a distance of $8$~kpc for all three sources,
and adopt the extinction of surrounding field stars as the foreground extinction to each
source, then we derive stellar masses of $12\pm3M_\odot$, $14\pm3M_\odot$,
$17\pm6M_\odot$ for SSTGC~524665, SSTGC~797384, and SSTGC~803187, respectively, by using
a grid of YSO models \citep{robitaille:06,robitaille:07}. More detailed discussion of
the model fitting will presented in a future paper.

To summarize, we presented the evidence from IRS spectra for the first spectroscopic
identification of massive YSOs in the GC. In our next paper (D.\ An, 2009, in preparation),
we will present the results for all 107 YSO candidates, together with additional data
from millimeter to radio observations, and use them to better understand the nature of
these embedded sources.

\acknowledgements

We thank David Ardila for helpful discussions of the IRS data reduction. We thank
Sean Carey for providing us MIPS photometry before publication. D.\ An and
S.\ Ram\'irez thank John Stauffer for helpful discussions.
This work is based on observations made with the Spitzer Space Telescope, which is
operated by the Jet Propulsion Laboratory, California Institute of Technology under
a contract with NASA. Support for this work was provided by NASA through an award
issued by JPL/Caltech.
This research has made use of the SIMBAD database, operated at CDS, Strasbourg, France.

\end{document}